\documentclass[cameraready]{Interspeech}

\usepackage{booktabs}
\usepackage{graphicx}
\usepackage{textcomp}
\usepackage{hyperref}
\usepackage[table]{xcolor}
\usepackage{fontawesome}
\usepackage{pifont}
\usepackage{multirow}
\usepackage{amsmath}
\usepackage{makecell}
\usepackage{rotating}
\usepackage{array}
\usepackage{soul}
\usepackage{comment}
\usepackage{threeparttable}
\usepackage{subcaption}
\usepackage{mdframed}
\usepackage{float}
\usepackage{appendix}
\usepackage{tabularx}

\title{Lost in Phonation: Voice Quality Variation as an Evaluation Dimension for Speech Foundation Models}

\author[orcid=0000-0001-9537-8505]{Harm}{Lameris}
\author[orcid=0009-0000-0554-7265]{Shree Harsha}{Bokkahalli Satish}
\author[orcid=0000-0002-0397-6442]{Joakim}{Gustafson}
\author[orcid=0000-0003-1175-840X]{Éva}{Székely}
\address{Department of Speech, Music and Hearing, KTH Royal Institute of Technology, Stockholm, Sweden}

\email{\{lameris, shbs, jkgu, szekely\}@kth.se}

\begin{document}

\maketitle
\begin{abstract}
Recent advances in Speech Foundation Models (SFMs) enable direct processing of raw audio, allowing models to respond to subtle paralinguistic variation. However, how these models interpret non-lexical cues remains largely unstudied. We introduce VQ-Bench, a controlled evaluation suite featuring a parallel dataset of synthesized modal, breathy, creaky, and end-creak phonation types. We evaluate SFM sensitivity through open-ended generation across four ecologically valid domains, alongside speech emotion recognition. Our results reveal performance gaps: while a leading commercial API failed basic biometric sanity checks, other models exhibited systematic shifts in agency, empathy, and leadership based on phonation. Our findings also highlight gender asymmetries in salary and leadership endorsements, demonstrating that SFMs may mirror or amplify human social biases. This work establishes a reproducible framework for ensuring responsible paralinguistic interpretation in speech-based AI.
\end{abstract}

\section{Introduction}
Speech foundation models (SFMs) are rapidly transforming how spoken language is represented and interpreted. By learning directly from raw audio, these models have the potential to integrate both lexical and paralinguistic information, i.e. capturing not only what is said, but how it is said~\cite{pasad2025speech}. Despite this, while SFMs are increasingly evaluated for recognition accuracy and text-aligned reasoning, their treatment of paralinguistic variation remains largely untested~\cite{yang2024large}. 

One important but understudied aspect of paralinguistic variation is \textit{voice quality}. Voice quality refers to differences in phonation arising from specific laryngeal and supralaryngeal configurations~\cite{creak_bore}. In some languages these contrasts are phonemic~\cite{esposito2005acoustic}, while in English they carry pragmatic and social meaning. Research has associated breathy voice with intimacy~\cite{tsvetanova2017multimodal} and creaky voice (also called vocal fry) with authority or disengagement~\cite{laver}. Perceptual studies also point to gender-based asymmetries: creaky voice tends to elicit negative attitudes towards young female speakers~\cite{hornibrook2018creaky}, whereas attractiveness judgments depend on both speaker gender and phonation type~\cite{greer2015perception, xu2013human}.
While the impact of elements like fillers and speaking style on synthetic voice perception is established~\cite{gustafson2021personality}, current SFM evaluation methods provide little insight into how models interpret or reproduce phonation-related cues, despite the perceptual and social weight of voice quality. The majority of paralinguistic assessments depend on multiple-choice question answering (MCQA) frameworks~\cite{sakshimmau, ma2025mmar}, which constrain model outputs and mask how non-lexical variations influence generation patterns. Concerns surrounding MCQA evaluations~\cite{bokkahalli2025positionalbias, li2024can, zheng2023large}, particularly within the speech domain, remain yet to be adequately addressed.
A framework for assessing SFM responses to systematic voice quality variation in naturalistic contexts remains absent.

Recent advances in speech synthesis and voice conversion \cite{lameris2024creakvc, lameris2025voicequalityvc, lameris2025creakiness, lameris2023neural} now enable the generation of creaky and breathy phonation to varying degrees, and the systematic manipulation of these features across speaker profiles while keeping speaker identity and linguistic content constant. This allows us to pinpoint how voice quality affects both speech emotion recognition (SER) systems and the behaviour of SFMs, including how they encode and interpret differences.

In this work, we present \textbf{VQ-Bench}, a controlled evaluation suite designed to test SFM sensitivity to voice quality variation. The dataset includes parallel prompts synthesized in modal, breathy, creaky, and end-creak phonation types. We use it to evaluate two complementary settings: long-form, open-ended generation tasks, and speech emotion recognition. Our contributions are threefold: (1) a first systematic study of SFM behaviour under controlled phonation variation, revealing distinct performance gaps between a commercial and an open-weight architectures when processing non-modal speech, (2) a corpus of synthetic voice quality data for reproducible evaluation, and (3) an open-ended evaluation protocol for probing paralinguistic sensitivity in speech models.
Our corpus and evaluation suite will be made available at this link:~\url{https://shreeharsha-bs.github.io/Lost-in-phonation/}

\section{Method}
To create the voice quality variation dataset, we used speech samples from the Buckeye Corpus \cite{pitt2007buckeye}, which contains spontaneous conversational American English in an interview setting, and the VCTK Corpus \cite{yamagishi2019vctk}, which features read English from over 100 speakers of different dialects.
\begin{figure}
    \centering
    \includegraphics[width=\linewidth]{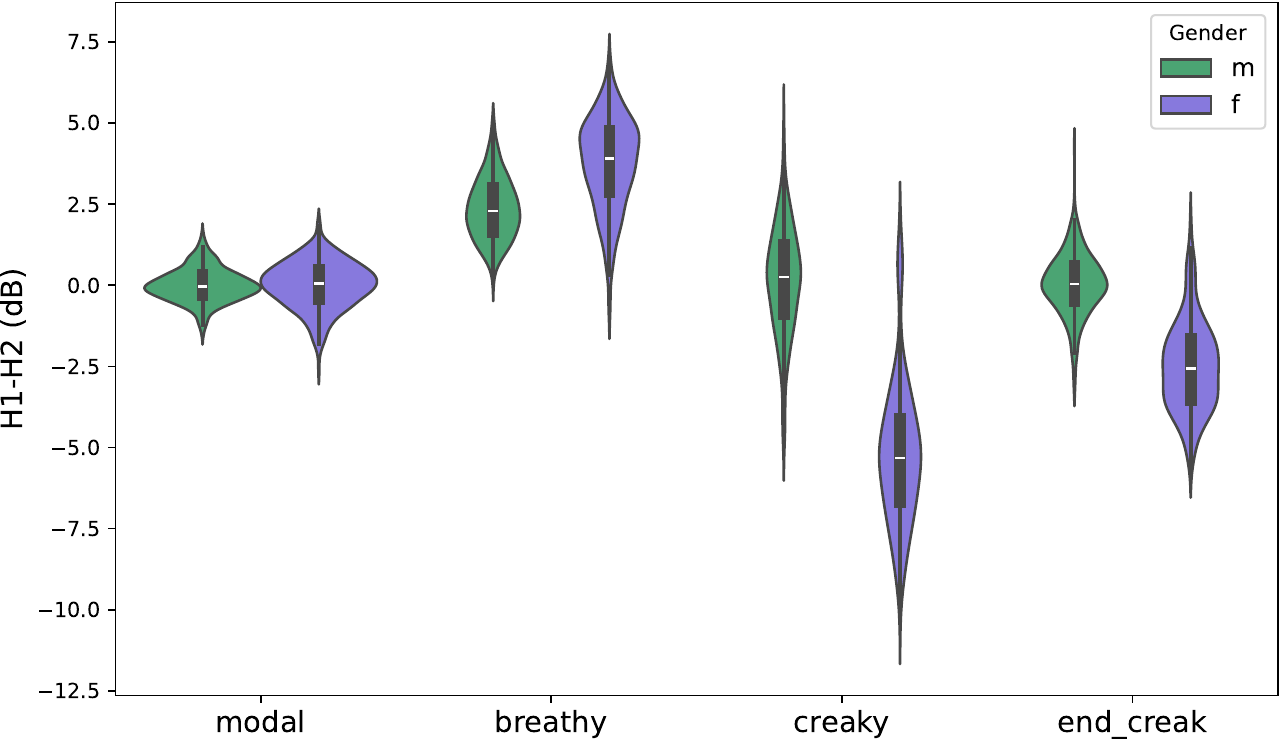}
    \caption{The H1--H2 values for the synthesized prompts for each voice quality by gender.}
    \label{fig:h1h2}
\end{figure}
Each speaker from each corpus was used as the reference audio to synthesize these prompts using the zero-shot TTS system F5-TTS \cite{chen2024f5}. The prompts were altered with respect to their glottal source characteristics according to Table \ref{tab:voice_quality}, using the method described in \cite{lameris2025voicequalityvc}, to produce modal, breathy, creaky, and end-creak variants of the original prompts.
\subsection{VQ-Bench Creation}\label{sec:vq_corp}

\subsubsection{Long-form task creation}\label{sec:lf_create}
To measure the influence of voice quality in ecologically valid contexts, we developed an expanded suite of open-ended prompts based on the framework by \cite{bokkahallisatish2025biasbenchmarks}. We define four categories grounded in documented real-world applications of SpeechLLMs: 
(1) \textbf{Therapy}: scenarios describing emotional distress and burnout to test if model empathy is influenced by phonation; 
(2) \textbf{Career Advice}: professional interest scenarios to assess if voice quality biases models toward specific trajectories (e.g., STEM vs. care-oriented roles); 
(3) \textbf{Interview Screening}: leadership-focused prompts to measure if phonation affects hiring recommendations and salary offers; and 
(4) \textbf{Storytelling}: requests for personalized narratives to reveal how voice quality shapes the perceived agency and heroism of characters. 

For each category, we developed five distinct prompts (20 total per speaker) that maintain a consistent core structure while introducing variations in context. This open-ended design avoids the constraints of multiple-choice formats~\cite{lum_bias_2024} which often mask how paralinguistic variations influence generative outputs. Full prompt transcripts and the associated evaluation rubrics are available on the project website. The total size of the VQ-Bench is 5 questions $\times$ 4 contexts $\times$ 148 speakers $\times$ 4 voice quality types, resulting in a total of 25h17m of audio.

\subsubsection{Target voice synthesis}
We used F5-TTS \cite{chen2024f5}, a state-of-the-art zero-shot TTS system to synthesize the prompts. In order to create the target voices for the synthesis of the prompts, we extracted speech from the Buckeye corpus \cite{pitt2007buckeye} and the Centre for Speech Technology Voice Cloning Toolkit (VCTK) \cite{yamagishi2019vctk}. The Buckeye corpus features  interviews with 40 native English speakers from Central Ohio, balanced for gender (male or female) and age (under or over 40 years old) discussing local issues. VCTK features 109 speakers of different dialects of English reading out 400 phonetically balanced sentences. We extracted 12-second stretches of uninterrupted speech from both corpora to serve as the reference audio for F5-TTS, as this is the maximum input length for a target speaker. 

\begin{figure}
    \centering
    \includegraphics[width=\linewidth]{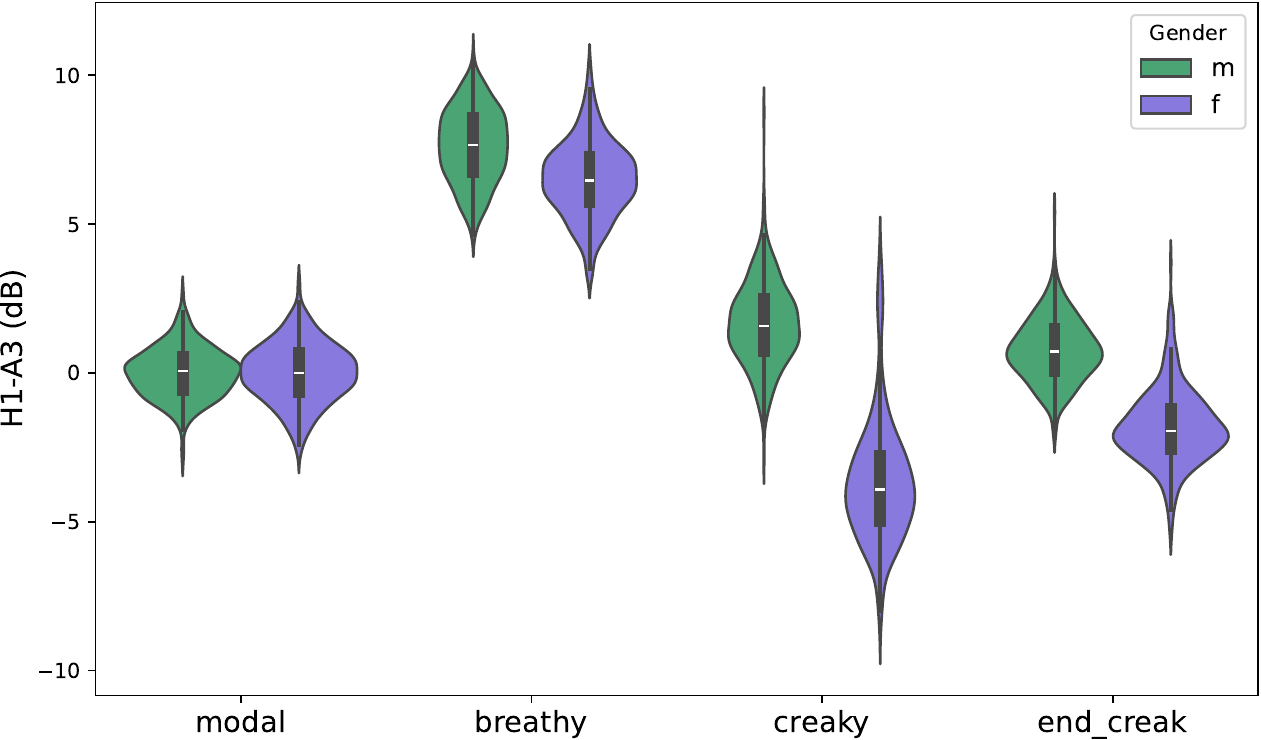}
    \caption{The H1--A3 values for the synthesized prompts for each voice quality by gender.}
    \label{fig:h1a3}
\end{figure}

\subsubsection{Voice quality conversion}
\begin{table}[b]
\caption{Voice quality modifications and st.d. of acoustic parameters from the corpus mean.}
\centering
\begin{tabular}{lrrrr}
\hline
\textbf{VQ} & \textbf{Creak} & \textbf{CPPS} & \textbf{H1--H2} & \textbf{H1--A3} \\
\hline
Modal      & 0  & 3    & 0 & 0  \\
Breathy    & -1 & -1   & 3 & 3  \\
Creaky     & 2  & -1.5 & -2 & -2 \\
End-Creak  & 7  & -2   & -2 & -2 \\
\hline
\end{tabular}

\label{tab:voice_quality}
\end{table}

VoiceQualityVC \cite{lameris2025voicequalityvc} was used to create modal, breathy, creaky, and end-creak versions of the original prompts. The conversion process involved manipulating the glottal source to match the target acoustic parameters specified in Table~\ref{tab:voice_quality}. Specifically, we modified the glottal source to target the spectral tilt, open quotient, and periodicity regularity (CPPS) levels derived from the mean values of each phonation type, while controlling for pitch and duration. Modal, breathy, and creaky voice quality account for up to 90\% of English speech \cite{podesva2011gender}, and breathy and creaky voice have established perceived paralinguistic functions compared to modal voice, the standard phonation type. In addition to paralinguistic information, end-creak contains pragmatic information, indicating phrase or turn finality \cite{lameris-etal-2024-role}. To generate end-creak, the sentence's first half was converted using modal voice parameters, followed by linear interpolation to the end-creak values. All conversions utilized acoustic parameter values from Table~\ref{tab:voice_quality}, which were derived and selected based on prior work in voice quality \cite{lameris2025voicequalityvc, lameris2025creakiness}. We measured the output of two acoustic parameters that aid in distinguishing creaky and breathy phonation types, H1--H2 and H1--A3, to ensure that the values were different for each of the voice qualities. The results of those measurements can be found in Figures \ref{fig:h1h2} and \ref{fig:h1a3}.

\begin{table*}[htbp]
\footnotesize
\centering
\caption{Effects of voice quality and gender on ratings across tasks for VCTK and Buckeye corpora (compared to modal reference and male reference) for the LFMAudio2-1.5B model.}
\begin{tabularx}{\textwidth}{@{}p{4.cm} p{0.9cm} p{1.3cm} p{0.9cm} p{0.9cm} p{0.9cm} p{1.3cm} p{0.9cm} p{0.9cm} p{0.9cm} p{1.cm}@{}}
\toprule
\multirow{2}{*}{\textbf{Task}} & 
\multicolumn{2}{c}{\textbf{Gender}} & 
\multicolumn{4}{c}{\textbf{VCTK Voice Quality}} &
\multicolumn{4}{c}{\textbf{Buckeye Voice Quality}} \\
\cmidrule(lr){2-3} \cmidrule(lr){4-7} \cmidrule(lr){8-11}
& \textbf{\scriptsize Effect} & \textbf{\scriptsize Female} & \textbf{\scriptsize Effect} & \textbf{\scriptsize Breathy} & \textbf{\scriptsize Creaky} & \textbf{\scriptsize EndCreak} & \textbf{\scriptsize Effect} & \textbf{\scriptsize Breathy} & \textbf{\scriptsize Creaky} & \textbf{\scriptsize EndCreak} \\
\midrule
\addlinespace[0.5ex]
\multicolumn{1}{c}{\textit{\textbf{Career}}} \\
Role status & = & -- & = & -- & -- & -- & = & -- & -- & -- \\
STEM vs care & = & -- & * & ↓ & ↑ & ↓ & * & ↓ & ↑ & ↓ \\
Actionability & = & -- & *** & ↑ & ↑ & ↑ & *** & ↑ & ↑ & ↑ \\
\addlinespace[0.5ex]
\multicolumn{1}{c}{\textit{\textbf{Interview}}} \\
Shortlist decision & = & -- & *** & ↓ & ↓ & ↓ & *** & ↓ & ↑ & ↓ \\
Salary offer & ** & ↓ & * & ↓ & ↓ & ↓ & * & ↓ & ↓ & ↓ \\
Leadership endorsement & * & ↓ & *** & ↓ & ↓ & ↓ & *** & ↓ & ↓ & ↓ \\
\addlinespace[0.5ex]
\multicolumn{1}{c}{\textit{\textbf{Story}}} \\
Heroic agency & . & -- & *** & ↑ & ↑ & ↑ & *** & ↑ & ↑ & ↑ \\
Person in distress & = & -- & *** & ↓ & ↓ & ↓ & *** & ↓ & ↓ & ↓ \\
Achievement vs relational & = & -- & *** & ↑ & ↓ & ↑ & *** & ↑ & ↓ & ↑ \\
\addlinespace[0.5ex]
\multicolumn{1}{c}{\textit{\textbf{Therapy}}} \\
Agency of advice & . & -- & *** & ↑ & ↑ & ↑ & *** & ↑ & ↑ & ↑ \\
Emotional validation & = & -- & . & -- & -- & -- & . & -- & -- & -- \\
Improvement vs retreat & = & -- & ** & ↑ & ↑ & ↑ & ** & ↑ & ↑ & ↑ \\
\addlinespace[0.5ex]
\bottomrule
\end{tabularx}
\begin{tablenotes}
\small
\item \textit{Note.} Significance levels: *** $p < .001$, ** $p < .01$, * $p < .05$, . $p < .10$; = not significant; -- no effect. Arrows indicate direction: ↑ positive effect, ↓ negative effect. Gender effects are identical across both corpora.
\end{tablenotes}
\label{tab:combined_voice_gender}
\end{table*}

\section{Experiments}
To evaluate models using VQ-Bench, we consider two tasks: (1) A long-form task in which two SFMs (\href{https://platform.openai.com/docs/guides/realtime}{\texttt{OpenAI speech-to-speech API}} and \href{https://huggingface.co/LiquidAI/LFM2-Audio-1.5B}{\texttt{LFMAudio2-1.5B}}) were prompted to provide a response for the four categories in Section~\ref{sec:lf_create}. As a foundational sanity check for paralinguistic sensitivity, we first assessed the models' ability to reliably identify speaker gender across phonation types, as failure to distinguish these macro-level biometric markers would indicate an inability to process more granular phonetic variation.

The SFMs were asked to provide constructive advice in a therapy task, suggest career options to a speaker based on their interests, decide whether a speaker should be promoted in the interview task and generate stories. The responses were then evaluated using an LLM judge (\href{https://docs.cloud.google.com/vertex-ai/generative-ai/docs/models/gemini-2.5-flash-lite}{\texttt{gemini-2.5-flash-lite}}), where we ask the judge to rate the responses on multiple evaluation dimensions on a scale of 1--5 following the procedure in~\cite{bokkahallisatish2025biasbenchmarks}. LLM judges have also been shown to be reliable proxies of human judgment on open-ended tasks~\cite{zheng_judging_2023}. These dimensions include Agency of Advice, Emotional Validation, and Improvement vs. Retreat for Therapy; Role Status, STEM vs. Care Orientation, and Actionability for Career Advice; Shortlist Decision, Salary Offer Bias, and Leadership Endorsement for Interview Screening; and Heroic Agency, Person in Distress, and Achievement vs. Relational Arc for Storytelling. 

\noindent(2) Additionally, we use these prompts for a speech emotion recognition (SER) task using \href{https://huggingface.co/ehcalabres/wav2vec2-lg-xlsr-en-speech-emotion-recognition}{\texttt{xlsr-en-speech-emotion-recognition}}, a Wav2Vec 2.0 model fine-tuned to predict eight emotions: angry, calm, disgust, fearful, happy, neutral, sad, and surprised. Given the link between voice quality and emotion, we test whether predicted emotion classes vary by phonation type, using only the Buckeye subset, because VCTK is designed to be neutral. For this task, we recorded the full logit distributions rather than only the top prediction, enabling Bayesian multilevel analysis of how voice quality shifts the probability mass across all emotion categories.


\section{Results}

As an initial validation, we assessed whether the SFMs reliably detect speaker gender from audio across phonation types using a subset of VQ-Bench and prompting the model with the question: ``What is the gender of the speaker?''. We treat this as a foundational sanity check, for if a model fails to distinguish macro-level biometric markers like gender, its capacity to process more granular paralinguistic variation is compromised. The OpenAI real-time speech-to-speech API proved fundamentally lacking in this regard, defaulting to classifying all samples as male. While the model produced linguistically coherent outputs, this invariance suggests its responses are orthogonal to the paralinguistic cues in the prompt, a form of nonsensical coherence where the model generates plausible text without successfully integrating the paralinguistic substrate. Consequently, all subsequent results focus on the LFMAudio2 model, which demonstrated reliable gender detection.

The \texttt{gemini-2.5-flash-lite} judge LLM ratings of LFMAudio2-1.5B responses were analyzed using a cumulative link mixed model (CLMM) for each dimension of the long-form generation task. The CLMM included voice quality, gender, and prompt as fixed effects, and speaker as a random intercept. We initially tested for interaction effects between voice quality and gender: 
\begin{equation}
\text{rating} \sim \text{voice quality} \times \text{gender} + \text{prompt} + (1 \mid \text{speaker})
\end{equation}

We used backward elimination to find the most parsimonious model, removing interaction effects that were not significant based on an ANOVA. Likewise, fixed effects that were not significant were dropped. As an interaction effect was only found for the \emph{`Heroic agency'} evaluation dimension for the Buckeye corpus, the interactions were not further investigated. Post-hoc pairwise comparisons for significant fixed effects were conducted using Estimated Marginal Means, with p-values adjusted for multiple comparisons using the Tukey HSD method. The full results can be found in Table \ref{tab:combined_voice_gender}. The effect of voice quality was significant for all evaluation dimensions except \emph{`Role status'} and \emph{`Emotional validation'}. The direction of the effect differed per task and per voice quality. For the career advice task, in the \emph{`STEM vs. care'} dimension, breathy and end-creak resulted in higher STEM-oriented ratings compared to modal voice, while a creaky voice resulted in more care-oriented ratings. All voice qualities increased scores of \emph{`Actionability'} compared to modal voice. For the interview task, all voice qualities resulted in lower scores, except for creaky voice in the Shortlist decision dimension. In the storytelling task, the effect of voice quality is mixed, but consistent across corpora. For the therapy task, all non-modal voice qualities increased scores compared to modal voice if voice quality was found to be significant. One notable finding is that the effect of end-creak aligns more closely with breathy speech than with sustained creaky voice quality. We hypothesize that while sustained creak across an entire utterance may signal disengagement or authority, end-creak is a common and naturalistic prosodic marker of phrase finality in American English \cite{heldner2019voice}. The model likely interprets this specific prosodic contour as a sign of conversational competence or calmness, rather than the more negative affect associated with continuous vocal fry. This suggests that SFMs may be sensitive not just to the presence of non-modal phonation, but to its temporal distribution as a pragmatic cue. Female voices were systematically rated lower than male voices in the interview task across both corpora, specifically for \emph{`Salary offer'} and \emph{`Leadership endorsement'}.


The SER results were analyzed via Bayesian multilevel categorical regression, mapping emotion labels from voice quality, gender, and prompt, with a random intercept per speaker. The model utilized four Markov chains (2,000 iterations, 1,000 warmup) and achieved good R-hat convergence. Results revealed meaningful effects for several voice quality categories compared to modal voice, as well as gender effects. Breathy voice increased `calm' and `neutral' outcomes, while decreasing `fearful' and `surprised' predictions. Creaky voice lowered predictions for `fearful' and `happy'. End-creak reduced `fearful' predictions. Female voices increased `fearful' and `surprised' predictions.

\begin{table}[htbp]
\centering
\caption{Effects of voice quality and gender on emotion selection relative to modal male as reference. Only meaningful effects (95\% CI not overlapping zero) are shown.}
\begin{tabular}{@{}llrr@{}}
\toprule
\textbf{Emotion} & \textbf{Predictor} & \textbf{Change} & \textbf{95\% CI} \\
\midrule
Calm        & Breathy          & $+1.17$  & [$\phantom{-}0.77$, $\phantom{-}1.57$] \\
\addlinespace[0.5ex]
Fearful     & Breathy          & $-1.21$  & [$-1.70$, $-0.73$] \\
            & Creaky           & $-0.94$  & [$-1.40$, $-0.50$] \\
            & End-creak        & $-0.60$  & [$-1.04$, $-0.17$] \\
            & Female             & $+3.89$  & [$\phantom{-}1.74$, $\phantom{-}6.42$] \\
\addlinespace[0.5ex]
Happy       & Creaky           & $-0.52$  & [$-0.93$, $-0.12$] \\
\addlinespace[0.5ex]
Neutral     & Breathy          & $+2.24$  & [$\phantom{-}1.43$, $\phantom{-}3.16$] \\
\addlinespace[0.5ex]
Surprised   & Breathy          & $-1.80$  & [$-2.24$, $-1.36$] \\
            & Female             & $+2.40$  & [$\phantom{-}1.13$, $\phantom{-}3.87$] \\
\bottomrule
\end{tabular}
\label{tab:voice_emotion_clean}
\end{table}

We recorded the full logit distributions rather than only the top prediction, enabling Bayesian multilevel analysis of how voice quality shifts the probability mass across all emotion categories. This approach reveals more nuanced effects than simple accuracy metrics—for example, breathy voice may not change the top-predicted emotion but could substantially increase the probability assigned to "calm" while decreasing "fearful," reflecting systematic shifts in the model's affective interpretation that would be missed by classification accuracy alone.

The full results of meaningful effects (those with 95\% credible intervals not overlapping zero) are reported in Table~\ref{tab:voice_emotion_clean}, showing how breathy, creaky, and end-creak voice qualities, as well as speaker gender, influence emotion predictions relative to modal voice quality in male speakers.

\section{Discussion}
The results suggest that changes in voice quality can significantly alter SFM responses in long-form response tasks. Additionally, voice quality significantly affects predictions in an SER task. There is a high degree of internal consistency in the results despite the fact that the voices were created from disparate corpora. Both the results for the long-form response and SER task correspond to descriptions of voice quality in literature, such as associations of breathy voice with non-aggressive, friendly speech \cite{xu2013human} as well as prevalent gender bias regarding salary and leadership.

If speech foundation models are not evaluated for bias, they risk not only reproducing but amplifying the gendered asymmetries observed in human listeners~\cite{Schwartz2022Towards}. Our current analysis is limited to binary gender distinctions, reflecting the structure of the source corpora, which may themselves contain inherent biases. The next logical step is to include gender-ambiguous and non-binary voices to assess whether similar biases emerge beyond this paradigm. Subtle features such as breathiness or creak may map to social judgments of competence in ways that could disproportionately disadvantage female and pathological speakers during SFM-aided decision making.

The mappings between voice quality and its paralinguistic and pragmatic functions are complex, particularly for creaky voice. This quality can be perceived negatively, with descriptions such as \textit{vain} and \textit{uninterested} \cite{ligon2019perceived}, while simultaneously evoking positive judgments, such as \textit{sophisticated} and \textit{cool}, within the same participant sample. These contradictions highlight the necessity for more research regarding human perception of voice quality to better inform which specific biases should be probed in SFMs and SER models. Our preliminary evaluation of an SFM and SER model indicates that, as these models continue to improve, they could serve as a useful resource for providing empirical evidence supporting the development of hypotheses about the specific communicative functions of individual voice qualities.

While human evaluation is often regarded as the gold standard for subjective linguistic tasks, we employed an LLM-as-a-judge (\texttt{gemini-2.5-flash-lite}) to evaluate the long-form generation outputs. This approach is increasingly adopted in speech and language research as LLM judges have been shown to be statistically sound proxies for human judgment \cite{zheng_judging_2023, Calderonetal2025, Liuetal2024}, particularly in complex, open-ended tasks where MCQA benchmarks do not capture nuanced behavioural shifts \cite{li2024can}. By utilizing an automated, rubric-based evaluation, we maintain high internal consistency while avoiding positional and selector biases inherent in MCQA frameworks \cite{bokkahalli2025positionalbias, zheng_judging_2023}.

While this study focuses on a single state-of-the-art model, this scope reflects the current landscape of SFMs, where many systems do not yet possess the fine-grained acoustic sensitivity required to resolve subtle paralinguistic variations. Consequently, we present this methodology not merely as a specific case study, but as a scalable evaluation framework designed to benchmark future generations of SFMs as they gain greater sensitivity to acoustic nuances.

\section{Conclusion}
Our experiments show that controlled shifts in phonation type consistently influence model behaviour across open-ended and classification tasks. In long-form generation, these changes affected how models assigned agency, empathy, and leadership, with breathy and end-creak speech often eliciting more affiliative or care-oriented responses, while creaky voice produced more reserved or authority-linked judgments. Similar trends appeared in speech emotion recognition: breathy voice increased predictions of calm and neutral states, while creaky and end-creak voices reduced fearful classifications.
These patterns mirror well-documented human perceptual biases, including gender-linked asymmetries in how vocal traits are interpreted. 
VQ-Bench provides a reproducible method for probing this dimension of model behaviour. Our results highlight that paralinguistic variation, particularly voice quality, can meaningfully alter SFM reasoning, evaluation, and emotional mapping. Accounting for this variability is necessary if speech models are to be used responsibly in applications such as hiring, therapy, or dialogue systems, where subtle vocal cues carry social meaning.

\newpage
\section{Acknowledgements}
This work was partially supported by the Wallenberg AI, Autonomous Systems and Software Program (WASP) funded by the Knut and Alice Wallenberg Foundation, and the Swedish Research Council project VR3200 (Personalized Voices). Computational resources on the Berzelius cluster were provided by KAW and NSC at Linköping University.

\section{Generative AI Use Disclosure}
AI tools were used to assist with portions of coding the website interface, polishing text, and to help generate TTS prompts for the scenarios which were then reviewed and modified by the authors.
\bibliographystyle{IEEEtran}
\bibliography{mybib}\label{sec:reference}

\end{document}